\documentstyle[12pt]{article}

\begin{document}
\begin{titlepage}
 \renewcommand{\thefootnote}{\fnsymbol{footnote}}
    \begin{normalsize}
     \begin{flushright}
                 UT-692\\
                 November 1994 \\
                 hep-th/9411051
     \end{flushright}
    \end{normalsize}
    \begin{Large}
       \vspace{1cm}
       \begin{center}
         {\bf  Anyon Basis
               of $c=1$ Conformal Field Theory
         } \\
       \end{center}
    \end{Large}

  \vspace{10mm}
\begin{center}
\begin{Large}
           Satoshi Iso\footnote
           {E-mail address: iso@danjuro.phys.s.u-tokyo.ac.jp} \\
\end{Large}
         \vspace{2cm}
         {\it Department of Physics, University of Tokyo,} \\
               {\it Bunkyo-ku, Tokyo 113, Japan}\\

\vspace{15mm}

\end{center}
\begin{abstract}
\noindent
We study the $c=1$ conformal field theory of a free compactified
boson with
radius $r=\sqrt{\beta}$ ($\beta$ is an integer). The Fock space
of this boson is constructed in terms of  anyon vertex operators
and each state is labeled by an infinite set of pseudo-momenta
of filled particles in pseudo-Dirac sea. Wave function of multi anyon state
 is described by an
eigenfunction of the Calogero-Sutherland (CS) model.
The $c=1$ conformal field theory at $r=\sqrt{\beta}$ gives a
field theory of CS model.
This is a natural generalization of the  boson-fermion
correspondence in one dimension to boson-anyon correspondence.
There is also
an interesting duality between anyon with statistics
$\theta=\pi/\beta$ and particle with statistics $\theta=\beta \pi$.
\end{abstract}

\end{titlepage}
\vfil\eject
\newpage
\section{Introduction}
\setcounter{equation}{0}
Recently there is a renewal of interest in the one-dimensional
(1d) solvable model with inverse-square type interactions.
This class of models includes Calogero-Sutherland model
\cite{calogero,sutherland}, spin 1/2 generalization by
Haldane and Shastry \cite{haldane,shastry},
models with internal symmetry
\cite{hahaldane,kawakami,minpoly,hikami}, supersymmetric t-J
model \cite{kuramoto} or hierarchical generalization
\cite{hierarchy}. Haldane-Shastry model has been  most deeply
analyzed and it is understood that the elementary excitations
are described by spinons with semionic statistics
\cite{haldane2}.
The ground state of the model is called Gutzwiller
wave function, a good variational wave function for
the Heisenberg model.
Different from the Heisenberg model, the correlation
function of the
Haldane-Shastry model has no logarithmic correction
and it indicates absence of marginally irrelevant
operators in the theory.
This means that the Haldane-Shastry model is the fixed point
Hamiltonian of the Heisenberg model \cite{yangian},
which is  the Wess-Zumino-Witten model
of $SU(2)$ at level 1.
This was confirmed recently \cite{bernard,bouwknegt}.
Since other models with inverse-square interactions also
have no logarithmic corrections in their correlation
functions, it is natural to think that all the models of this type
are the fixed point Hamiltonian of interacting theories.
In particular,
there are many indications \cite{c=1} that
the Calogero-Sutherland (CS) model with coupling
constant $\beta$ on a circle
\begin{equation}
H_{cs} =
\sum_i {p_i^2 \over 2} + \left( {\pi \over L} \right)^2
\sum_{i<j} {\beta^2-\beta \over \sin^2 {\pi \over L}(x_i-x_j) }.
\label{CShamiltonian}
\end{equation}
will give the  $c=1$ conformal field theory of a
compactified boson with radius $r=\sqrt{\beta}$.
\par
The $c=1$ conformal field theory (CFT) of
a compactified boson has a rich structure \cite{ginzparg}.
At the special radius point $r=1$,
the bosonic Fock space is also described by free Dirac fermions.
At $r=\sqrt{2}$,
there  is an affine  $SU(2)$ symmetry and
Fock space is constructed by these generators.
At $r=\sqrt{3}$, there is an $N=2$ supersymmetry.
Among them, $r=1$ theory is special because, in fermionic
representation, each state is labeled by an infinite set of
momenta of filled particles in the Dirac sea.
For example, if we mode-expand fermion operator
$\psi^{\dagger}(z)$ by
\begin{equation}
\psi^{\dagger}(z)=\sum_{n \in {\bf Z}}
\psi^{\dagger}_{-n-1/2} z^{n},
\end{equation}
the charge $0$ ground state is written as
\begin{equation}
|0 \rangle = \psi^{\dagger}_{1/2} \psi^{\dagger}_{3/2}
    \cdot \cdot \cdot  | - \infty \rangle
    = \prod_{i=0}^{\infty} \psi^{\dagger}_{-m_i^{0}} | - \infty \rangle
    \label{fermionvacuum}
\end{equation}
 where $m_i^{0}=-(2i-1)/2$.
 In general, a state in the fermion basis is written as
 \begin{equation}
 | \{ m_i \} \rangle = \prod_{i=0}^{\infty} \psi^{\dagger}_{-m_i}
 | - \infty \rangle.
\end{equation}
$\{ m_i \} $ ( $m_i > m_j$ if $i<j$ ) are infinite set of momenta of
filled particles in the Dirac sea.
No such structure is known for other compactified bosons.
\par
In this paper, we show that, as well as
the free fermion theory, there is a
pseudo-momentum representation of states
at other radius points
$r=\sqrt{\beta}$, where $\beta$ is an integer.
At these points, the Fock space can be created by
anyon vertex operators. The pseudo-Dirac sea
is constructed by fermion or boson vertex operators
corresponding to odd $\beta$ or even $\beta$.
Each state is labeled by an infinite set of integers
$m_i$ ($i=1,2,...$) where $m_1>m_2> ...$ but,
different from the free fermion, there is a restriction
$|m_i - m_j| \geq \beta$
to the value of pseudo-momenta.
We mainly consider a case with integer $\beta$
but it is straightforward to generalize the result to
a case with rational $\beta=p/q$.
\par
The paper is organized as follows.
In section 2, we give a brief review of Jack symmetric
functions and the Calogero-Sutherland model.
In section 3, we first construct $c=1$ CFT Fock space by
multi-anyons (Section 3.3) and then by {\it super fermion}
vertex operators (Sec.3.4). By using these construction,
we give a  pseudo-Dirac sea description of the Fock space (Sec.3.5).
We also give a collective Hamiltonian (Sec.3.6).
Section 4 is devoted to conclusion and discussions.
\section{Symmetric Functions and  Calogero-Sutherland Model}
\setcounter{equation}{0}
\subsection{Jack Polynomials}
First we  give a brief review of Jack symmetric functions.
Here we use the notation of Macdonald \cite{Mac}.
Let $(x_1,..,x_N)$ be independent indeterminates.
There are many basis of symmetric functions for these valuables,
each of which are characterized by a {\it partition}, or
Young tableau. It is a sequence
\begin{equation}
\{ \lambda \} = (\lambda_1, \lambda_2, ...)
\end{equation}
of non-negative integers, such that
$\lambda_1 \geq \lambda_2 \geq \cdot \cdot \cdot$.
The degree of the partition is defined by
\begin{equation}
|\lambda|= \sum \lambda_i.
\end{equation}
The conjugate partition $\{ \lambda' \}$ is defined to interchange rows and
columns in the corresponding Young tableau.
The nonzero $\lambda_i$ are called the parts of the partition
and the number of the parts is the length $l(\lambda)$ of the
partition. If $\{ \lambda \}$ has $m_1$ parts equal to $1$ ,
$m_2$ parts equal to $2$, and so on, the partition is
written as $\{ \lambda \}=(1^{m_1} 2^{m_2} ...)$.
To this partition, we define
\begin{equation}
z_{\lambda} \equiv \prod_r r^{m_r} m_r!
\end{equation}
The {\it partial ordering} is defined in the set of all  partitions
as follows:
\begin{equation}
\{ \lambda \} \geq \{ \mu \} \Longleftrightarrow
|\lambda|=|\mu| \ \ \  {\mbox and} \ \ \
\sum_i^r \lambda_i \geq \sum_i^r \mu_i \ \ \ {\mbox for \ \  all} \ r \geq 1.
\end{equation}
For each partition we define the following basis of symmetric functions;
\par
(1) Monomial symmetric functions \\
To each partition $\{ \lambda \}$ monomials $x_{\{\lambda \} }$
are defined by $x_{ \{ \lambda \} }=x_1^{\lambda_1}x_2^{\lambda_2} ...$.
Monomial symmetric functions $m_{\{\lambda \} }$
are obtained from $x_{\{ \lambda \} }$
by permutations of the $x's$.
\par
(2) Power sums \\
$N$-th power sum is defined by $p_n=\sum_i x_i^n$.
For each partition, $p_{\{ \lambda \} }$ is defined by
$p_{\{ \lambda \} }= p_1^{\lambda_1} p_2^{\lambda_2} ...$. \\
\par
Now we define the Jack symmetric functions.
Let's define a scalar product
\footnote{Note that $\alpha$ in the Macdonald's notation \cite{Mac}
or in the Stanley's notation \cite{stanley} is $\alpha= 1/\beta$.}
on a space of symmetric functions
\begin{equation}
\langle p_{ \{ \lambda \} }, p_{\{ \mu \} } \rangle =
\delta_{\{ \lambda \},\{ \mu \}} \beta^{-l(\lambda)} z_{\lambda}.
\label{innerproduct}
\end{equation}
Then one can define the Jack symmetric functions
$J_{\{ \lambda \} }^{\beta}(x_1, ...x_N)$
by the following
conditions:
\begin{eqnarray}
&(a)& J_{\{ \lambda \} }^{\beta}(x_i)= m_{ \{ \lambda \} } +
\sum_{ \{ \mu \} < \{ \lambda \} } v_{\{ \lambda \},\{ \mu \}}(\beta)
m_{ \{ \mu \} }
\nonumber \\
&(b)&
\langle J_{\{ \lambda \} }^{\beta}, J_{\{ \mu \} }^{\beta} \rangle = 0,
\ \ \ {\mbox if} \ \ \     \{ \lambda \} \neq  \{ \mu \}.
\end{eqnarray}
$v_{\{ \lambda \},\{ \mu \}}(\beta)$ is a coefficient.
For $\beta=1$ the Jack symmetric function coincides with the Schur
functions for free fermions.
The norm of the Jack symmetric functions is defined by
\begin{equation}
\langle J_{\{ \lambda \} }^{\beta}, J_{\{ \lambda \} }^{\beta} \rangle
=  j_{\{ \lambda \} }^{(\beta)}.
\end{equation}
In particular, $j_{\{ 0 \} }=1$.
\par
The Jack symmetric functions have a very interesting property of
{\it duality} $\beta \longleftrightarrow 1/\beta$.
Let $\omega$ be an automorphism on the ring of symmetric polynomials,
defined by
\begin{equation}
\omega(p_n)=- {(-1)^n  p_n \over \beta}.
\end{equation}
Then duality transformation transforms a Jack polynomial  into its dual:
\begin{equation}
\omega J_{\{ \lambda \} }^{\beta} =
j_{\{ \lambda \} }^{(\beta)} J_{\{ \lambda' \} }^{(1/\beta)}
\end{equation}
Since power sums form a basis of the ring of symmetric functions,
Jack symmetric functions can be written as a functional of
power sums $p_n$;
\begin{equation}
J_{\{ \lambda \} }^{\beta}(x_1, x_2,...x_N)=
J_{\{ \lambda \} }^{\beta}(\{ p_n \}).
\end{equation}
With this notation, the above duality tells us that
\begin{equation}
J_{\{ \lambda \} }^{(\beta)}( \{- {p_n \over \beta} \} ) =
(-1)^{|\lambda|} j_{\{ \lambda \} }^{(\beta)}
J_{\{ \lambda' \} }^{(1/\beta)}( \{p_n \}).
\label{dualtrans}
\end{equation}
Two dual Jack symmetric functions are related by the following
duality relation:
\begin{equation}
\prod_{i=1}^N \prod_{j=1}^M
(1 - x_i y_j) = \sum_{\{ \lambda \} } (-1)^{|\lambda|}
J_{\{ \lambda \} }^{(\beta)}(x_i) J_{\{ \lambda' \} }^{(1/\beta)}(y_j)
\label{duality}
\end{equation}
Here $\{ \lambda' \}$ is a conjugate partition to $\{ \lambda \}$
and partitions $\{ \lambda \}$ are summed over those that satisfy
$l(\lambda) \leq N$ and $l(\lambda') \leq M$.
\subsection{Calogero-Sutherland Model}
Jack symmetric functions introduced above give  eigenfunctions of the
Calogero-Sutherland (CS) model, an exactly solvable model in 1d with
inverse-square interactions on a circle. The Hamiltonian is
given by (\ref{CShamiltonian}).
Using $z$ variables instead of $x$, defined by
$z=e^{i2 \pi x/L}$, the Hamiltonian is rewritten as
\begin{equation}
H_{CS}= {1 \over 2} \left( {2 \pi \over L } \right)^2
\left\{   \sum(z_i \partial_i)^2 - 2\beta(\beta-1)
\sum_{i<j} {z_i z_j \over (z_i - z_j)^2 }  \right\}
\label{CSham2}
\end{equation}
where $\partial_i = \partial/\partial z_i$. The groundstate has the Jastraw
form:
\begin{equation}
\psi_0(z_i)= \prod_{i<j} (z_i - z_j)^{\beta} \prod_i z_i^{-\beta N/2}.
\label{ground}
\end{equation}
($N$ is a particle number.)
Wave function of  an excited state is  written as
$ \psi(z_i)=f(z_i)\psi_0(z_i) $ where $f(z_i)$ is a symmetric polynomial.
The CS Hamiltonian (\ref{CSham2}) acts on $f(z_i)$ as
\begin{eqnarray}
 && H_{CS} \psi(z_i) =({\tilde H}_{CS} f(z_i)) \psi_0, \ \ \
 {\tilde H}_{CS} = {\tilde H} +
 {1 \over 2} \left( {2 \pi \over L } \right)^2
 {N^3 - N \over 12} \beta^2
\nonumber \\
\label{gaugetr}
&& \tilde{H} ={1 \over 2} \left( {2 \pi \over L } \right)^2
\left\{   \sum(z_i \partial_i)^2 + \beta
\sum_{i<j} {z_i + z_j \over z_i - z_j } (z_i \partial_i -z_j \partial_j)
 \right\}.
\nonumber \\
\end{eqnarray}
Its eigenfunction is characterized by a Young tableau $ \{ \lambda \}$
\cite{forrester}:
\begin{equation}
f(z_i)= J_{\{ \lambda \}}^{(\beta)}(z_i) \prod_i^N (z_i)^p
\end{equation}
whose eigenvalue of ${\tilde H}_{CS}$ is
\begin{eqnarray}
&& {1 \over 2} \left( {2 \pi \over L } \right)^2
\sum_i^N  (m_i)^2
\nonumber \\
 && m_i = p + \lambda_i +\beta ({N+1 \over 2}-i).
 \end{eqnarray}
 When we write $p=q-\beta N /2 $,
 \begin{equation}
 m_i = q+ \lambda_i - {2 i-1 \over 2} \beta.
 \end{equation}
 The above state is also an eigenstate of the momentum operator
 \begin{equation}
 {\hat P} =  {2 \pi \over L} \sum z_i \partial_i
 \end{equation}
 with an eigenvalue  $ \sum_i^N m_i $.
In later sections, this momentum operator is identified with the
Virasoro generator while the CS Hamiltonian is identified with
a higher-spin (spin 3) conserved charge.
\subsection{Collective Hamiltonian}
The CS Hamiltonian ${\tilde H}$ acts on a space of symmetric functions
and since the ring of symmetric functions are generated by power sums
$p_n=\sum z_i^n$,
we can rewrite the CS Hamiltonian  in terms of creation and
annihilation of power sums.
By using the standard technique of collective field theory
\cite{coll1,coll2,coll3}, we can obtain the collective Hamiltonian
of the CS model. $f$ can be expanded in terms of the power sums as
\begin{equation}
f=\sum_k f\{n_1, ... n_k \} \prod_{i=1}^k p_{n_i} \equiv
f\{n_1, ... n_k \} |n_1, ... n_k \rangle
\end{equation}
where $f \{ n_i \}$ are expansion coefficients. Annihilation and
creation operators of
power sums ($n>0$)
are defined by $|\{ n \} \rangle \equiv a_n^{+} |\{ 0 \} \rangle$
and $a_n |\{ n  \} \rangle = n |\{0  \} \rangle$.
They satisfy the commutation relation
$[a_n, a_m^{+}]= n \delta_{m,n}$ and the "vacuum" conditions
$a_n|\{ 0  \} \rangle =0$.
Note that they are not hermite conjugate in general:
$(a_n)^{\dagger} \neq a_n^{+} $.
In terms of $a_n$ and $a_n^{+}$, we get the collective
field theory Hamiltonian for the CS Hamiltonian ${\tilde H}$:
\begin{equation}
{\tilde H}_{coll} = \sum_{n>0} (1-\beta) n a_n^+ a_n +N \beta \sum_{n>0}
a_n^+ a_n + \sum_{n,n'>0} (\beta a_n^{+} a_{n'}^{+} a_{n+n'}
+ a_{n+n'}^{+} a_n a_{n'} ).
\label{collective1}
\end{equation}
The eigenstates are given  by
\begin{equation}
J_{\{ \lambda \}}^{(\beta)}(\{ a_n^{+} \}) |\{ 0 \} \rangle.
\label{collective2}
\end{equation}
The second term in the Hamiltonian contains the particle number,
but as we will see later, this $N$ dependence can be absorbed in the
last term.
\section{$c=1$ CFT}
\setcounter{equation}{0}
\subsection{Vertex Operators}
In this section we consider a compactified free boson with radius
$r=\sqrt{\beta}$.
We only consider the holomorphic part here.
$\beta$ can take any positive number but when we consider
the statistics of vertex operators or the structure of
pseudo-Dirac sea, we restrict it to an integer.  It is straightforward
to generalize it to a rational number $\beta=p/q$.
\par
First the bosonic field is expanded as
\begin{equation}
\phi(z) = {\hat q}-i \alpha_0 \log z + i \sum_{n \neq 0}
{\alpha_n z^{-n} \over n}.
\end{equation}
Commutation relations are $[{\hat q}, \alpha_0] =1$ and
$[\alpha_n,\alpha_{n'}^{\dagger}]=n \delta_{n,n'}$.
The bosonic field satisfies the operator product expansion (OPE):
\begin{equation}
\phi(z) \phi(z')  \sim  - \log(z-z').
\end{equation}
Energy-momentum tensor is defined by
$T(z) \equiv  -(1/2):(\partial \phi(z))^2:$
and we  define the charge by the zero mode of the current operator
$J(z) \equiv i \sqrt{\beta} \partial \phi(z)$, that is,
$\sqrt{\beta} \alpha_0$.
The Fock space of this compactified boson is characterized by the
charge $q$
and bosonic excitations constructed by $\alpha_n$ for $n<0$.
The charge must take  integer values.
\par
We can now define two kinds of vertex operators. One is anyon vertex
operators and the other is  bosonic or fermionic vertex operators
corresponding to even $\beta$ or odd $\beta$.
Anyon vertex operators are defined by
\begin{equation}
\Phi^{\pm}(z) = :e^{\pm i \phi(z) / \sqrt{\beta}}:.
\end{equation}
The scaling dimension of the operators is  $1/(2 \beta)$ and they
have anyonic statistics $\theta= \pm \pi / \beta$.
They carry  charge $\pm 1$.
OPE's are given by
\begin{eqnarray}
\Phi^{\pm}(z) \Phi^{\pm}(z') &=&
       (z-z')^{ 1/\beta} :e^{\pm i(\phi(z)+\phi(z'))/\sqrt{\beta} }:
      \sim (z-z')^{1/\beta} \nonumber \\
\Phi^{+}(z) \Phi^{-}(z') &=&
       (z-z')^{- 1/\beta} :e^{i(\phi(z)-\phi(z'))/\beta}:
       \nonumber \\
     &\sim & (z-z')^{-1/\beta}(1+ (z-z')
        {i \over \sqrt{\beta} } \partial \phi(z')  + ...).
\end{eqnarray}
Mode expansion on a charge $q$ sector is given by
\begin{equation}
\Phi^{\pm}(z) |q \rangle = \sum_{n \in {\bf Z}}
\Phi^{\pm}_{-n \mp {q \over \beta} -{1 \over 2 \beta} } z^{n \pm {q \over
\beta}}
                |q \rangle
\end{equation}
By following the standard procedure \cite{zam} we can derive the following
generalized commutation relations
\footnote{They are derived by \cite{bouwknegt} for $\beta=2$ and used
to obtain a fermionic representation of Virasoro characters
and $SU(2)$ level 1 Kac-Moody algebra.}
:
\begin{eqnarray}
&& \sum_{l=0}^{\infty} C_l^{1-1/\beta} (
\Phi^{\pm}_{1-n_1 \mp {q \pm 1 \over \beta}-l-{1 \over 2\beta}}
\Phi^{\pm}_{-n_2 \mp {q \over \beta}+l-{1 \over 2\beta}} +
(n_1 \longleftrightarrow n_2)    )   |q \rangle =0
\nonumber \\
&& \sum_{l=0}^{\infty} C_l^{-1+1/\beta} (
\Phi^{+}_{-1 -n_1 - {q-1 \over \beta}-l-{1 \over 2\beta}}
\Phi^{-}_{-n_2 + {q \over \beta}+l-{1 \over 2\beta}}
\nonumber \\
  && \ \ \ \ \ +
\Phi^{-}_{-1 -n_2 +{q+1 \over \beta}-l-{1 \over 2\beta}}
\Phi^{+}_{-n_1 - {q \over \beta}+l-{1 \over 2\beta}} ) |q \rangle
= \delta_{n_1+n_2,-1}|q \rangle.
\end{eqnarray}
The coefficients are defined by
$ (1-x)^{\alpha}= \sum_{l=0}^{\infty} C_l^{\alpha} x^l$.
When $\beta=1$, these commutation relations reduce to the usual
fermion anti-commutation relations.
\par
The other kind of vertex operators are
\begin{equation}
\Psi^{\pm}(z) = :e^{\pm i \sqrt{\beta} \phi(z) }:.
\end{equation}
They have scaling dimension $\beta/2$ and statistics $\pm \pi \beta$.
Let's call this kind of vertex operator {\it super fermion} vertex operator
since, if $\beta$ is an integer, it has an integer-valued (times $\pi$)
exchange phase as the usual free bosons or free fermions.
They carry  charge $\pm \beta $.
OPE's are given by
\begin{eqnarray}
\Psi^{\pm}(z) \Psi^{\pm}(z') &=&
       (z-z')^{\beta} :e^{\pm i \sqrt{\beta} (\phi(z)+\phi(z'))}:
      \sim (z-z')^{\beta} \nonumber \\
\Psi^{+}(z) \Psi^{-}(z') &=&
       (z-z')^{- \beta} :e^{i \sqrt{\beta} (\phi(z)-\phi(z'))}:
       \nonumber \\
     &\sim & (z-z')^{-\beta}(1+ (z-z')
        i \sqrt{\beta}  \partial \phi(z')  + ...).
\end{eqnarray}
Mode expansions on a charge $q$ sector are similarly given by
\begin{equation}
\Psi^{\pm}(z) |q \rangle = \sum_{n \in {\bf Z}}
\Psi^{\pm}_{-n \mp q  -{\beta \over 2 } } z^{n \pm q }
                |q \rangle.
\end{equation}
In this case, the charge $q$ can be absorbed by  shift of $n$.
Generalized commutation relations
for $\Psi$ and $\Psi$ or $\Phi$ and $\Psi$
are similarly given.
Here we only comment
 that there is a simple relation for $\Psi$'s:
\begin{equation}
\Psi^{\pm}_{-n_1-{\beta \over 2}} \Psi^{\pm}_{-n_2-{\beta \over 2}}
-(-1)^{\beta}
\Psi^{\pm}_{-n_2-{\beta \over 2}} \Psi^{\pm}_{-n_1-{\beta \over 2}}
=0
\end{equation}
\subsection{Pseudo-Dirac Sea --- Vacuum}
By using the above mode expansion, we can define the pseudo-Dirac sea
with an infinite set of pseudo-momenta.
First
it is easy to check that the vacuum state for the boson annihilation operators
with charge $q$ satisfies the following conditions:
\begin{eqnarray}
 && \Phi^{\pm}_{n \mp {q \over \beta}-{1 \over 2 \beta} } |q \rangle =0 ,
\ \ \ n \geq 1
 \nonumber \\
 && \Psi^{\pm}_{n \mp q  -{\beta \over 2 } } |q \rangle =0,
 \ \ \ n \geq 1.
 \end{eqnarray}
Similar to the free fermion case eq.(\ref{fermionvacuum}), we can
express charge $0$ vacuum in terms of an infinite set of pseudo-momenta
as follows:
\begin{equation}
|0 \rangle = \Psi_{\beta \over 2}^{\dagger} |-\beta \rangle=
\Psi_{\beta \over 2}^{\dagger} \Psi_{3 \beta \over 2}^{\dagger}
\cdot \cdot \cdot  | - \infty \rangle
    = \prod_{i=0}^{\infty} \Psi^{+}_{-m_i^{0}} | - \infty \rangle
\end{equation}
where  $m_i^{0}=-(2i-1) \beta /2$.  Different from the free fermion case,
however, commutation relations between $\Psi^{+}$ and $\Psi^{-}$ are
 complicated and such states as
$\prod_{i=0}^{\infty} \Psi^{+}_{-m_i}
 | - \infty \rangle$
 are not orthogonal to each other in general.
 We must therefore find more general framework to construct pseudo-momentum
 representation or pseudo-Dirac sea. As we can see later,
 there are two trivial cases (other than vacuum)
 that the above way of construction of states works.
 One is a state that can be created by a single anyonic vertex operator
 $\Phi^{+}$
 and the other is a state that can be created by a single
 super-fermionic vertex operator
 $\Psi^{+}$. In the language of Young tableau, these states correspond
 to tableau that has only single row or single column.
\subsection{Multi Anyon States}
In order to obtain a general framework for pseudo-Dirac sea, we
first consider multi anyon states. The duality relation (\ref{duality})
plays an important role. The left hand side of the equation
can be rewritten as
\begin{eqnarray}
\prod(1-x_i y_j) &=& \exp(\log \prod(1-x_i y_j) )=
\exp [-\sum_{n=1}^{\infty} \sum_{i,j} { (x_i y_j)^n \over n} ]
\nonumber \\
&=& \exp [-\sum_{n=1}^{\infty} \sum_{j=1}^{M} { p_n(x) (y_i)^n \over n}].
\end{eqnarray}
Here $p_n(x)=\sum_i^N (x_i)^n$.
Therefore, the duality relation (\ref{duality}) becomes
\begin{equation}
\exp [-\sum_{n=1}^{\infty} \sum_{j=1}^{M} { p_n(x) (y_j)^n \over n}]
= \sum_{\{ \lambda \} } (-1)^{|\lambda|}
J_{\{ \lambda \} }^{(\beta)}(\{ p_n \} )
J_{\{ \lambda'\} }^{(1/\beta)}(y_j).
\label{anyon1}
\end{equation}
The sums are restricted to $l(\lambda) \leq N$ and $l(\lambda') \leq M$.
If we put formally $N \rightarrow \infty$ and
$p_n(x)=\mp \alpha_{-n} / \sqrt{\beta}$, we get
\begin{equation}
\exp [\pm   \sum_{n=1}^{\infty} \sum_{j=1}^{M}
 { \alpha_{-n}\over \sqrt{\beta}}  {(y_j)^n \over n}]
= \sum_{\{ \lambda \} } (-1)^{|\lambda|}
J_{\{ \lambda'\} }^{(1/\beta)}(y_j)
J_{\{ \lambda \} }^{(\beta)}(\{ \mp {\alpha_{-n} \over \sqrt{\beta}} \} ).
\end{equation}
Sums are restricted to $l(\lambda') \leq M$.
\par
$M$-anyon states on the vacuum with charge ${\tilde q}$ are obtained by
multiplying $M$ anyon vertex operators:
\begin{eqnarray}
&& \prod_{j=1}^{M} \Phi^{\pm}(z_j) |{\tilde q} \rangle =
\prod_{i<j} (z_i -z_j)^{1/\beta} :e^{\pm i \sum \phi(z_j) /\sqrt{\beta} }:
|{\tilde q} \rangle  \nonumber \\
&&= \prod_{i<j} (z_i -z_j)^{1/\beta} \prod_j^M  (z_j)^{\pm {\tilde q} /\beta}
\sum_{\{ \lambda \} } (-1)^{|\lambda|}
J_{\{ \lambda'\} }^{(1/\beta)}(z_j)
J_{\{ \lambda \} }^{(\beta)}(\{ \mp {\alpha_{-n} \over \sqrt{\beta}} \} )
|\pm M+{\tilde q} \rangle \nonumber \\
&&= \prod_{i<j} (z_i -z_j)^{1/\beta} \prod_j^M  (z_j)^{\pm {\tilde q} /\beta}
\sum_{\{ \lambda \} } (-1)^{|\lambda|}
J_{\{ \lambda'\} }^{(1/\beta)}(z_j) \  ( j_{\lambda}^{\beta})^{1/2}
 | \{ \lambda \}, \pm M+{\tilde q} \rangle_{\mp}.
 \nonumber \\
 \label{multianyon}
\end{eqnarray}
Here we defined
\begin{equation}
| \{ \lambda \}, q \rangle_{\mp} \equiv
(j_{\lambda}^{\beta})^{-1/2}
J_{\{ \lambda \} }^{(\beta)}(\{ \mp {\alpha_{-n} \over \sqrt{\beta}} \} )
|q \rangle.
\label{lambdastate}
\end{equation}
At $\beta=2$ this state was identified with the motif-represented state
of $SU(2)$ level 1 Kac-Moody algebra \cite{bernard}.
These states are shown to be orthogonal to each other
and normalized to unity.
This can be proved as follows:
If we define {\it power-sum} states corresponding to
a partition $\{ \lambda \} $, created by the bosonic creation
operators $\mp \alpha_{-n}  \sqrt{\beta}$,
\begin{equation}
| \alpha_{\{ \lambda \} }, q \rangle_{\mp}
\equiv
\prod_{i=1}^{l(\lambda)} \left(
\mp {\alpha_{\lambda_i}^{\dagger} \over \sqrt{\beta}} \right)
|q \rangle,
\end{equation}
it is easy to prove that these states satisfy
\begin{equation}
 _\mp \langle   \alpha_{\{ \lambda \} }, q |
 \alpha_{\{ \mu \} }, q \rangle_{\mp}
 = \delta_{\{ \lambda \},\{ \mu \}} \beta^{-l(\lambda)} z_{\lambda}.
\end{equation}
Therefore, from the definition of the Jack symmetric functions,
those states
$| \{ \lambda \}, q \rangle_{\mp} $ defined above are orthogonal
to each other and normalized to unity.
Moreover, due to the completeness of Jack symmetric functions,
they form a complete basis in the $c=1$ CFT
if we consider all partitions and
all integer charges.
\par
{}From the duality
transformation eq.(\ref{dualtrans}),
these states can be written in another form:
\begin{equation}
| \{ \lambda \}, q \rangle_{\mp} =
(-1)^{|\lambda|}
(j_{\lambda}^{\beta})^{1/2}
J_{\{ \lambda' \} }^{(1/\beta)}(\{ \pm  \sqrt{\beta} \alpha_{-n}  \} )
|q \rangle.
\label{statelambda}
\end{equation}
\par
For a partition $\{ \lambda \}$ where $l(\lambda') \leq M$,
from eq.(\ref{multianyon}) and orthonormal property of
$| \{ \lambda \}, q \rangle_{\mp}$, we get
\begin{equation}
 _\mp \langle \{ \lambda \},\pm M + {\tilde q}|
\prod_{j=1}^{M} \Phi^{\pm}(z_j) |{\tilde q}  \rangle =
\prod_{i<j} (z_i -z_j)^{1/\beta} \prod_j^M  (z_j)^{\pm {\tilde q} /\beta}
 (-1)^{|\lambda|}
 J_{\{ \lambda'\} }^{(1/\beta)}(z_j) \  ( j_{\lambda}^{\beta})^{1/2}
 \label{betainvwave}
\end{equation}
The r.h.s is the eigenfunction of the CS Hamiltonian with coupling
constant $1/\beta$.
This shows that the eigenfunction of $1/\beta$ CS model
gives "bound state" wave functions for $M$ anyon states.
\subsection{Super Fermion Picture}
Next let's consider states created by {\it super fermion} vertex
operators $\Psi^{\pm}$. Similar to the multi anyon case,
action of multi super-fermion vertex operators can be evaluated
as follows. First note that eq.(\ref{anyon1}) can be also written in the
following form:
\begin{equation}
\exp [-\sum_{n=1}^{\infty} \sum_{i=1}^{N} { p_n(y) (x_i)^n \over n}]
= \sum_{\{ \lambda \} } (-1)^{|\lambda|}
J_{\{ \lambda \} }^{(\beta)}( x_i)
J_{\{ \lambda'\} }^{(1/\beta)}(\{ p_n \}).
\end{equation}
Then put formally $M \rightarrow \infty$ and
$p_n(y)=\mp \sqrt{\beta} \alpha_{-n}$ and we get
\begin{equation}
\exp [\pm \sum_{n=1}^{\infty} \sum_{i=1}^{N}
 \sqrt{\beta} { \alpha_{-n} (x_i)^n \over n}]
= \sum_{\{ \lambda \} } (-1)^{|\lambda|}
J_{\{ \lambda \} }^{(\beta)}( x_i)
J_{\{ \lambda'\} }^{(1/\beta)}(\{ \mp \sqrt{\beta} \alpha_{-n}  \}).
\end{equation}
Sums are restricted to $l(\lambda) \leq N$.
Using this equation and  eq.(\ref{statelambda}), action of
$N$ super-fermion vertex operators on vacuum is obtained:
\begin{eqnarray}
&& \prod_{i=1}^{N} \Psi^{\pm}(z_i) |{\tilde q} \rangle =
\prod_{i<j} (z_i -z_j)^{\beta} :e^{\pm i \sqrt{\beta} \sum \phi(z_j)  }:
|{\tilde q} \rangle  \nonumber \\
&&= \prod_{i<j} (z_i -z_j)^{\beta} \prod_i^N  (z_i)^{\pm {\tilde q} }
\sum_{\{ \lambda \} } (-1)^{|\lambda|}
J_{\{ \lambda\} }^{(\beta)}(z_i)
J_{\{ \lambda'\} }^{(1/\beta)}(\{ \mp \sqrt{\beta} \alpha_{-n}  \} )
|\pm N \beta +{\tilde q} \rangle \nonumber \\
&&= \prod_{i<j} (z_i -z_j)^{\beta} \prod_i^N  (z_i)^{\pm {\tilde q}}
\sum_{\{ \lambda \} }
J_{\{ \lambda \} }^{(\beta)}(z_i) \  ( j_{\lambda}^{\beta})^{-1/2}
 | \{ \lambda \}, \pm N \beta +{\tilde q} \rangle_{\pm}.
 \nonumber \\
 \label{multifermion}
\end{eqnarray}
Summations are over partitions $l(\lambda) \leq N$.
Therefore, for a partition $l(\lambda) \leq N$, bound state wave functions
of multi super-fermion states are given by
\begin{equation}
 _\pm \langle \{ \lambda \}, \pm q|
\prod_{i=1}^{N} \Psi^{\pm}(z_i) |\pm(q- N \beta) \rangle =
\prod_{i<j} (z_i -z_j)^{\beta} \prod_i^N  (z_i)^{q- N \beta }
J_{\{ \lambda \} }^{(\beta)}(z_1,...z_N) \  ( j_{\lambda}^{\beta})^{-1/2}
\label{betawave}
\end{equation}
Here we redefined the charge so that the bra-state has charge $\pm q$.
The r.h.s. is the eigenfunctions of the CS Hamiltonian with coupling
constant $\beta$. This shows that the  super-fermion
vertex operator can be interpreted as a field operator of interacting
fermions of CS model. The  $c=1$ CFT at $r=\sqrt{\beta}$, therefore, gives a
second quantized formulation of CS model with coupling $\beta$.
Furthermore, eq.(\ref{betainvwave}) and eq.(\ref{betawave}) show that there are
dual picture of $c=1$ CFT, "anyon" picture and  "super-fermion"
picture. In the anyon picture, multi-anyon bound states are
characterized by $1/\beta$ CS eigenfunctions. On the other hand,
in the super fermionic picture, multi fermion bound states are
characterized by $\beta$ CS eigenfunctions.
\subsection{Pseudo-Momentum Representation}
Now we get the framework to study  the pseudo-Dirac sea structure
of the $c=1$ CFT. The r.h.s. of eq.(\ref{betawave}) is an
eigenfunction of the
CS Hamiltonian $H_{CS}$ with an eigenvalue \begin{equation}
{1 \over 2} \left( {2 \pi \over L } \right)^2
\sum_i^N  (m_i)^2
\end{equation}
where $m_i$ is given by
\begin{equation}
m_i = q + \lambda_i -\beta {2i - 1 \over 2}.
 \end{equation}
 If we put formally $N \rightarrow \infty$,
it is characterized by an infinite set of
pseudo-momenta. Since the state $| \{ \lambda \}, \pm q \rangle_{\pm}$
is independent of $N$, we can label the state by this infinite set
of pseudo-momenta. For the vacuum state with charge $0$, this
infinite set $\{ m_i^0 \} $ coincides with  (\ref{fermionvacuum}).
Since $\lambda_1 > \lambda_2 > ...$,
the pseudo-momenta ($m_i > m_j$ if $i<j$)
can take any integer value
with the only constraints $|m_i - m_j| \geq \beta$.
{}From now on, we write
 \begin{equation}
 |\{ m_i \} \rangle_{\pm} \equiv | \{ \lambda \}, \pm q \rangle_{\pm}
 \label{pm-rep}
 \end{equation}
Each ($\pm$) pseudo-momenta representation of states
form an orthonormal and complete basis in the $c=1$ CFT.
\par
Here let's see how the Virasoro generator acts on the pseudo-momentum
representation. From the operator product expansion of the
Virasoro generator $T(z) \equiv  -(1/2):(\partial \phi(z))^2:$
and  the vertex operator $\Psi$, we get the following equation for an
analytic function $f(z)$:
\begin{eqnarray}
&& \langle \{ m_i \}| \oint {dz \over 2 \pi i} f(z) T(z) \prod \Psi^{+}(z_i)
|{\tilde p} \rangle =
\nonumber \\
&& \oint {dz \over 2 \pi i} f(z)
\sum_i  \left( {\beta/2 \over (z-z_i)^2} + {1 \over z-z_i}\partial_i
\right) \langle \{ m_i \}| \prod \Psi^{+}(z_i) |{\tilde p} \rangle.
\end{eqnarray}
Virasoro generators $L_n$ are given by putting $f(z)=z^{n+1}$.
If we put $f(z)=z$ and $N \rightarrow \infty$, we get
\begin{equation}
L_0  |\{ m_i \} \rangle_{\pm} = (\sum_i^{\infty} m_i)
|\{ m_i \} \rangle_{\pm}.
\label{infinitesum}
\end{equation}
On the other hand
the action of $L_0$ is also given by eq.(\ref{pm-rep}) and
eq.(\ref{lambdastate}) and  the eigenvalue of $L_0$ is
\begin{equation}
\sum \lambda_i + {q^2 \over 2 \beta}.
\end{equation}
Regularizing the infinite sum of $m_i$ in eq.(\ref{infinitesum}),
it coincides with this
 eigenvalue. The second term is the Casimir energy of the charge
$q$ vacuum.
\subsection{Collective Hamiltonian}
In this subsection we give a collective Hamiltonian acting on
the state $|\{ m_i \} \rangle_{\pm}$ that gives an eigenvalue
$\sum_i^{\infty}  (m_i)^2 $. From eq.(\ref{pm-rep}), eq.(\ref{lambdastate}) and
eq.(\ref{collective2}), such collective Hamiltonian for states
$|\{ m_i \} \rangle_{\pm} $ is obtained by
replacing $a_n^{+}$ in (\ref{collective1})
by $\mp \sqrt{\beta} \alpha_n^{\dagger}$.
{}From commutation relations,  $a_n$ must be replaced by
$\mp \alpha_n  \sqrt{\beta}$.
Then we get the collective Hamiltonian
\begin{equation}
{\tilde H}_{coll}^{\beta, \pm} = \sum_{n>0} (1-\beta) n \alpha_n^{\dagger}
\alpha_n +N \beta \sum_{n>0} \alpha_n^{\dagger} \alpha_n \mp \sqrt{\beta}
\sum_{n,n'>0}  ( \alpha_n^{\dagger}\alpha_{n'}^{\dagger} \alpha_{n+n'}  +
\alpha_{n+n'}^{\dagger} \alpha_n \alpha_{n'} ).
\end{equation}
Comparing with eq.(\ref{ground}) and eq.(\ref{betawave}),
this collective Hamiltonian should be understood as acting on a state
with charge $\pm q$ that satisfies $q-N \beta = -N \beta /2$.
 Hence, for a general charged state,
the collective Hamiltonian is obtained by replacing
$N$ by $\alpha_0$ through the
relation $\alpha_0 \sqrt{\beta} =\pm q=\pm N \beta /2$:
\begin{equation}
{\tilde H}_{coll}^{\beta, \pm} =
\sum_{n>0} (1-\beta) n \alpha_n^{\dagger} \alpha_n
\mp \sqrt{\beta}
\sum_{n,n' \geq 0}  ( \alpha_n^{\dagger}\alpha_{n'}^{\dagger} \alpha_{n+n'}  +
\alpha_{n+n'}^{\dagger} \alpha_n \alpha_{n'} ).
\end{equation}
Of course, this Hamiltonian is hermite under the usual measure
for the free boson Fock space.
Eigenvalue of ${\tilde H}_{coll}^{\beta, \pm}$ on
$|\{ m_i \} \rangle_{\pm} $ is
\begin{equation}
{\tilde H}_{coll}^{\beta, \pm} |\{ m_i \} \rangle_{\pm} =
\sum_i^{\infty} ( (m_i)^2 -(m_i^0)^2 )  |\{ m_i \} \rangle_{\pm}.
\end{equation}
\par
The second term in the Hamiltonian is written in a local form
${1 \over 3} \oint {dz \over 2 \pi i} :(\partial \phi(z))^3: z^2$
but the first term cannot.
If we write the creation (annihilation) part of the bosonic field
by $\phi_{\pm}$, the collective Hamiltonian can be written as
\begin{equation}
{\tilde H}_{coll}^{\beta, \pm} =
\oint {dz \over 2 \pi i}      \left(  \mp
{\sqrt{\beta} \over 3} :(\partial \phi(z))^3: z^2 +
(1-\beta) :\partial \phi_+(z) (z\partial)^2\phi_-(z):   \right).
\end{equation}
\par
This collective Hamiltonian for the CS model is spin 3 conserved charge.
 This indicates that there are infinitely many conserved
charges $\sum_i^{\infty} ( (m_i)^p -(m_i^0)^p )$ in this $c=1$ CFT.
 Indeed it has been known that the Calogero-Sutherland model
is exactly solvable and has an infinitely many
conserved charges. In the paper \cite{conservedcharge}, these higher
conserved charges are discussed.
 Different from
the free fermion case, these charges will be non-local in general.
As the infinite charges in the free fermion obey $c=1$
$W_{1+\infty}$ algebra,
they will also obey some {\it deformed} $c=1$  $W_{1+\infty}$ algebra.
It will be interesting to find the full
 algebra and its representation theory.
\par
Eq.(\ref{betainvwave}) shows that there is another Hamiltonian that acts
on the pseudo-momentum states, CS Hamiltonian with coupling constant
$1/\beta$. Its collective Hamiltonian is given by replacing
$\beta$ by $1/\beta$ in ${\tilde H}_{coll}^{\beta, \pm}$.
This becomes
\begin{equation}
{\tilde H}_{coll}^{1/\beta, \pm}={-1 \over \beta}
      {\tilde H}_{coll}^{\beta, \mp}
\end{equation}
and gives the same collective Hamiltonian as that of $\beta$.
In other words, the same conserved charge in the $c=1$ CFT gives
different representation, CS Hamiltonian with coupling $\beta$
for multi super-fermion states and CS Hamiltonian with coupling $1/\beta$
for multi anyon states. This duality will be discussed in detail in a
separate paper.
\section{Conclusion and Discussions}
In this paper we study the $c=1$ conformal field theory with respect to
the multi anyon and  ``super-fermion''  states. By using the technique of
Calogero-Sutherland model, we show that at radius (of a compactified boson)
$r=\sqrt{\beta}$, where $\beta$ is an integer, the Fock space is labeled
by an infinite set of pseudo-momenta.
This gives a natural generalization of boson-fermion correspondence to
boson-anyon correspondence in (1+1) dimensions.
Bouwknegt et.al. \cite{bouwknegt} recently
obtained the Virasoro character at $r=\sqrt{2}$
point by using spinon description and Yangian symmetry.
It is an open problem to obtain the character
at other radius in this anyon basis.
\par
This anyon picture will be generalized to other
conformal field theories. For example, $c<1$ Virasoro minimal models
can be obtained from $c=1$ CFT by reducing the Fock space through
Feigin-Fucks construction. The anyon basis of $c=1$ CFT will give
a powerful tool to study the minimal models. This direction
is discussed by a paper \cite{matsuo}.
\par
Dynamical correlation functions for the
Calogero-Sutherland model have been recently obtained
\cite{dynamical,coll3}. Our formulation of the Calogero-Sutherland
model in terms of
$c=1$ CFT  will give a second quantized formulation of the dynamical
correlations. It will be published elsewhere.
\par
\subsection*{Acknowledgments}
I wish to thank T. Eguchi and K. Hori for valuable discussions and
useful comments. It is also a pleasure to acknowledge A. Kato, H. Kubo,
Y. Matsuo  and  J. Shiraishi for discussions.
This work is supported in part by
Grant-in-Aid for Scientific Research from Ministry of Science and
Culture.


\begin{thebibliography}{1}
\bibitem{calogero} F. Calogero, J. Math. Phys. {\bf 10} (1969) 2197;
  J. Math. Phys. {\bf 12} (1971) 418
\bibitem{sutherland}
B. Sutherland, J. Math. Phys. {\bf 12} (1971)246, 251; Phys. Rev. {\bf A4}
(1971) 2019; {\bf A5} (1971) 1372
\bibitem{haldane}
H. D. M. Haldane, Phys. Rev. Lett. {\bf 60} (1988) 635
\bibitem{shastry}
B. S. Shastry, Phys. Rev. Lett. {\bf 60} (1988) 639
\bibitem{hahaldane}
Z. N. C. Ha and H. D. M. Haldane,
Phys. Rev. {\bf B46} (1992) 9359
\bibitem{kawakami}
N. Kawakami, Phys. Rev. {\bf B46} (1992) 1005;
{\bf B46} (1992) 3191
\bibitem{minpoly}
J. A. Minahan and A. P. Polychronakos,
Phys. Lett. {\bf 302} (1993) 265
\bibitem{hikami}
K. Hikami and M. Wadati, Phys. Lett. {\bf A173} (1993) 263
\bibitem{kuramoto}
Y. Kuramoto and H. Yokoyama, Phys. Rev. Lett. {\bf 67} (1991) 1338
\bibitem{hierarchy}
N. Kawakami, Phys. Rev. Lett. {\bf 71} (1993) 275,
    J.Phys.Soc.Jpn {\bf 62}(1993) 2270, 2419  and cond-mat/9402011
\bibitem{haldane2}
  H. D. M. Haldane, Phys. Rev. Lett. {\bf 66} (1991) 1592
\bibitem{yangian}
  H. D. M. Haldane, Z. N. C. Ha, J. C. Talstra, D. Bernard
  and V. Pasquier,
  Phys. Rev. Lett. {\bf 69} (1992) 2021
\bibitem{bernard}
    D. Bernard, V. Pasquier and D. Servan, hep-th/9404050 (unpublished)
\bibitem{bouwknegt}
    P. Bouwknegt, A. W. W. Ludwig and K. Schoutens,
    hep-th/9406020 (unpublished)
\bibitem{c=1}
N. Kawakami and S-K, Yang,  Phys. Rev. Lett. {\bf 67} (1991)2493,
S. Iso and S. J. Rey, hep-th/9406192 (unpublished)
\bibitem{ginzparg}
  P. Ginsparg, Nucl. Phys. {\bf B 295}[FT 21] (1988) 153
\bibitem{Mac}
  I. G. Macdonald, Seminaire Lotharigien, Publ. I. R. M. A.
  Strasbourg, {\bf 372/S-20} (1988) 131
\bibitem{stanley}
  R. P. Stanley, Adv. in Math. {\bf 77} (1989) 76
\bibitem{forrester}
See, for example, P.J. Forrester, Nucl. Phys. {\bf B416} (1994) 377
\bibitem{coll1}
A. Jevicki and B. Sakita, Phys. Rev. {\bf D22} (1980) 467
\bibitem{coll2}
I. Andric, A. Jevicki and H. Levine, Nucl. Phys. {\bf B 215} (1983) 307
\bibitem{coll3}
J. A. Minahan and A. P. Polychronakos, hep-th/9404192 (unpublished)
\bibitem{zam}
A. B. Zamolodchikov and V. A. Fateev, Sov. Phys. J. E. T. P. {\bf 62}
(1985) 215
\bibitem{conservedcharge}
K. Hikami and M. Wadati, J. Phys. Soc. Jpn. {\bf 62} (1993) 4203
\bibitem{matsuo}
H. Awata, Y. Matsuo, S. Odake and J. Shiraishi,
hep-th/9411053  (unpublished)
\bibitem{dynamical}
F. D. M. Haldane and M. R. Zirnbauer,
Phys. Rev. Lett. {\bf 71} (1993) 4055, \
Z. N. C. Ha, Phys. Rev. Lett. {\bf 60} (1994) 1574, \
F. Lesage, V. Pasquier and D. Servan, Saclay preprint, (unpublished),
P. J. Forrester, cond-mat/9408042 (unpublished)
\end{thebibliography}
\end{document}